\documentclass{aastex62}
\usepackage{mathtools}

\shorttitle{Shallow Transits Deep Learning II.} 
\shortauthors{Dvash et al.}

\defcitealias{ZucGir2018}{Paper~I}

\begin{document}

\title{Shallow Transits --- Deep Learning II:\\
Identify Individual Exoplanetary Transits in Red Noise using Deep Learning}

\correspondingauthor{Shay Zucker}
\email{shayz@post.tau.ac.il}

\author{Elad Dvash}
\affiliation{Porter School of the Environment ans Earth Sciences, Raymond and Beverly Sackler Faculty of Exact Sciences, Tel Aviv University, Tel Aviv, 6997801, Israel}

\author{Yam Peleg}
\affiliation{Porter School of the Environment ans Earth Sciences, Raymond and Beverly Sackler Faculty of Exact Sciences, Tel Aviv University, Tel Aviv, 6997801, Israel}

\author[0000-0003-3173-3138]{Shay Zucker}
\affiliation{Porter School of the Environment ans Earth Sciences, Raymond and Beverly Sackler Faculty of Exact Sciences, Tel Aviv University, Tel Aviv, 6997801, Israel}

\author[0000-0002-2830-0297]{Raja Giryes}
\affiliation{School of Electrical Engineering, Iby and Aladar Fleischman 
Faculty of Engineering, Tel Aviv University, Tel Aviv, 6997801, Israel}

\begin{abstract}

In a previous paper, we have introduced a deep learning neural network that should be able to detect the existence of very shallow periodic planetary transits in the presence of red noise. The network in that feasibility study would not provide any further details about the detected transits. The current paper completes this missing part. We present a neural network that tags samples that were obtained during transits. This is essentially similar to the task of identifying the semantic context of each pixel in an image -- an important task in computer vision, called `semantic segmentation', which is often performed by deep neural networks. The neural network we present makes use of novel deep learning concepts such as U-Nets, Generative Adversarial Networks (GAN), and adversarial loss. The resulting segmentation should allow further studies of the light curves which are tagged as containing transits. This approach towards the detection and study of very shallow transits is bound to play a significant role in future space-based transit surveys such as \textit{PLATO}, which are specifically aimed to detect those extremely difficult cases of long-period shallow transits. Our segmentation network also adds to the growing toolbox of deep learning approaches which are being increasingly used in the study of exoplanets, but so far mainly for vetting transits, rather than their initial detection.

\end{abstract}


\section{Introduction} \label{sec:intro}

In a previous paper (\citealt{ZucGir2018}; hereafter \citetalias{ZucGir2018}), we have demonstrated a new approach to detect the presence of exoplanetary transits in simulated data mimicking data that can be obtained by high-cadence space telescopes. The demonstration was performed on simulated data of a fictitious telescope, but the approach should be applicable to real-life missions like \textit{CoRoT} \citep{Deletal2010}, \textit{Kepler} \citep{Boretal2010}, \textit{TESS} \citep{Ricetal2015}, and in the future \textit{PLATO} \citep{Rauetal2016}. The new approach we have introduced aimed at overcoming the problem of `red noise', usually attributed mainly to stellar activity, which constituted a major hurdle to traditional transit detection techniques, such as the BLS \citep{Kovetal2002}. Our suggested approach was based on the rapidly evolving new discipline of \textit{Deep Learning}. In \citetalias{ZucGir2018} we have demonstrated how this technique managed to outperform the BLS (preceded by a high-pass filter), in identifying light curves that contained exoplanetary transits, contaminated by red noise in addition to photon (Poisson) white noise.

It is important to note that \citetalias{ZucGir2018} focused on the task of detecting the presence of transits in the light curves, and not validating or vetting them as exoplanetary signals. The aim was to detect those transit events that might evade detection by traditional detection approaches like the BLS. In that respect it differed from other efforts in the field \citep[e.g.][]{ShaVan2018,Ansetal2018,Datetal2019,Liaetal2019,Osbetal2019}.

As successful as it may be, the detection mechanism we had introduced in \citetalias{ZucGir2018} lacked one crucial ingredient: it could not provide any information as to the details of the detected transits. The information it provided was binary: whether the light curve contained transits or not. Our aim in the current work is to present a deep learning neural network that will also identify the individual transits in the light curve, thus enabling further research, such as vetting the transit candidates, characterizing the transit properties, detecting transit timing variations (TTV), looking for additional transiting planets etc.

Deep learning is a class of algorithms and heuristics meant to train highly nonlinear parametric functions. The nonlinear functions, mostly known as neural networks, are essentially concatenations of layers of basic units, each comprising a linear operation followed by a simple nonlinearity. The nonlinearity is commonly realized by element-wise activation functions such as the sigmoid, hyperbolic tangent or the rectified linear unit (ReLU) \citep{NaiHin2010}. Their combination eventually results in intricate highly nonlinear functionality. 

During the training, the parameters of each layer are trained so as to minimize an error function calculated in relation to the previous layer. The training is often done using stochastic gradient descent \citep{Rumetal1986}. This approach often captures strong nonlinear relationships, leading to unprecedentedly successful results across many fields \citep[e.g.][]{Lecetal2015, Sch2015, Gooetal2016}

The task of identifying samples that are included in individual transits is essentially equivalent to the task of `semantic segmentation' in computer vision. The goal of the segmentation task is usually to simplify and change the representation of an image into something that is more meaningful and easier to analyze. Image segmentation is essentially the partitioning of a digital image into multiple segments, usually corresponding to objects and boundaries (lines, curves, etc.) in the image. Thus, image segmentation can be described as the process of assigning a label to every pixel in an image such that pixels with the same label share certain characteristics, or simply belong to the same context. The equivalence to the task of identifying the transits in a light curve is obvious: we assign a label to each sample in the lightcurve such that all the samples within transits get the same label. Since much progress has been achieved in performing image segmentation using deep learning, it is only natural to apply it here as well.

Most of the aforementioned studies, aimed at detecting, vetting, and identifying transits, made use of convolutional neural networks. In the current work, we use more tools from the toolkit of neural networks to perform segmentation. In particular, we use U-Nets \citep{Ronetal2015} to perform the segmentation and identify the times when a transit occurs within a given light curve signal, and an adversarial loss to force the network to output only realistic segmentation. 

In the next section we introduce the neural network concepts that we employed in our work. In Section~\ref{sec:impl} we present the way we implemented those concepts in our neural network. Section~\ref{sec:sim} describes the simulated dataset used for our demonstration, and the procedure we used to train the network is detailed in Section~\ref{sec:training}. Section~\ref{sec:results} demonstrates the performance of the neural network, and in Section~\ref{sec:discuss} we conclude and discuss the possible future implementations of the approach.

\section{Neural Networks} \label{sec:NN}

The approach we suggest for solving the problem of identifying individual transits makes use of several variants of deep learning neural networks: Convolutional Neural Networks (CNNs), ResNets, U-Nets, and Generative Adversarial Networks (GANs). In the next paragraphs we briefly introduce and explain these concepts, as well as other concepts we employ.

\subsection{Convolutional Neural Networks}\label{subsec:CNN}

In CNNs, convolutions constitute the linear part of the layers \citep{Lecetal1998}. CNNs are widely used to analyze images or periodic signals due to their shift-invariance property. CNNs are usually built by stacking convolution operators in layers, each followed by a non-linearity (`activation function'). Usually, the stack of convolutions is followed by a fully connected layer, represented by a simple linear function (matrix multiplication) followed by an activation function. These networks are known to be very powerful when applied to signal classification tasks (see \citetalias{ZucGir2018}). The layers are usually `contracting', in the sense that they perform successive downsampling of the signal, resulting in an increasingly compact representation of the information.

\subsection{Residual Networks (ResNets)}\label{subsec:ResNet}

An essential step in training neural networks is back-propagating the gradient of the loss function through the layers. A notorious problem in training very deep networks is the problem of `vanishing gradients': during the back-propagation of the gradients, repeated multiplications cause the gradients to become too small for effective learning. As a result, as networks grow deeper, the performance plateaus and might even start to degrade. A standard technique to avoid this problem uses \textit{residual connections} (also known as skip connections): retaining the original output of an earlier layer and adding it to the results of following layers as a `bypass' \citep{Heetal2016}. This helps to mitigate the vanishing gradient problem by causing the gradient from the earlier layer to flow through the bypass and skip multiplication steps.

\subsection{Fully Convolutional Networks and U-Nets}\label{subsec:UNet}

Building upon the concept of a CNN, a more elaborate deep learning architecture has emerged -- the \textit{fully convolutional network} (FCN), which is very popular for image segmentation \citep{Lonetal2015}. A popular FCN structure is the \textit{U-Net} (named after the U-shape of the network), which has been initially developed for biomedical images but has become widely used in many domains \citep[]{Ronetal2015}. Essentially, it is a CNN that is composed of two parts, the `encoder' and the `decoder'. The encoder is a contracting CNN, which produces a compact representation of the input signal. The decoder, which is appended to the encoder, comprises mirrored layers (with respect to the encoder), in the sense that each convolution in the encoder is mirrored by a corresponding \textit{deconvolution} (transposed convolution) layer. As a consequence, this expansive path is more or less symmetric to the contracting path, yielding a U-shaped architecture. It has been found that these networks perform better with the following improvement: depth-wise concatenation of the output of each encoding layer to the corresponding decoding layer in the mirrored architecture \citep[]{Ronetal2015}. By design, the original U-Net takes two-dimensional (2D), single-channel (gray scale) images as inputs. The current study deals with light curves, which are 1D (one-dimensional) time series. We have therefore restructured the U-Net design to take 1D time series as inputs by using 1D convolution layers. 

\subsection{Dice Loss}\label{subsec:Dice}

The objective of the training of a neural network is the minimization of a prescribed \textit{loss function}. The loss function represents the task one wishes the network to perform, and its goal is to provide a metric that measures the performance of the network for the given task. While in the problem of classification \citepalias[e.g.][]{ZucGir2018} the loss function commonly used is the logarithmic loss (also known as the cross-entropy loss), in this work we chose to apply a variant of the \textit{Dice Loss}, which is useful for segmentation problems.

Historically, the Dice coefficient was inspired by a set-theoretic concept introduced independently by \citet{Dic1945} and \citet{Sor1948} in ecological contexts in order to quantify similarity of sets. In the set-theoretic context, the Dice coefficient of the two sets $X$ and $Y$ is defined by:
\begin{equation}
\label{eq:dice_loss_sets}
    d(X,Y) = \frac{2|X \bigcap Y|}{|X| + |Y|}
\end{equation}
where $| \cdot |$ denotes the number of elements in each set. The concept of set membership is generalized to binary sequences, and the Dice coefficient for two binary sequences  $\{y_i\}$ and $\{p_i\}$ can now be written as:
\begin{equation}
\label{eq:dice_loss_general}
d(y,p) = \frac{2\sum_{i}p_{i}y_{i}}{{\sum_{i}p_{i}^2} + {\sum_{i}y_{i}^2} } .
\end{equation}

\citet{Miletal2016} were the first to apply the Dice coefficient to image segmentation. In this context, it is essentially a measure of the overlap between the segmentation image that the network produces and the ground-truth segmentation sequence. In our context, for a given light curve, let us denote the ground truth by a binary sequence $\{y_i\}$, where each sample in transit is assigned the value $1$ whereas all the rest are assigned $0$. Let $\{p_i\}$ denote the output of our segmentation network (the `prediction', in machine learning jargon), which is the probability (a value between $0$ and $1$) that sample $i$ is within a transit. We wish this probability to be $1$ for samples during transit and $0$ otherwise. Then the Dice coefficient for this light curve is given by:
\begin{equation}
\label{eq:dice_loss}
l = d(y,p) + d(1-y, 1-p),
\end{equation}
where $d(y,p)$ measures the performance during transit segments and $d(1-y,1-p)$ during out-of-transit segments.

\subsection{Adversarial loss}\label{subsec:adv_loss}

Our aim in the current project is to label samples occurring in transits. Naively, under the assumption that transits are strictly periodic (neglecting TTV and multiple transiting planets), it should have been very easy to judge whether the results of the segmentation are realistic. However, we wished to leave our mechanism agnostic of the strict periodicity of the signals, to allow, in future developments, the detection of multiplanetary signals, or signals with significant TTV. Thus, it is quite difficult to define a metric to measure how realistic is the resulting segmentation. Therefore, in training our neural network we make use of a novel concept in deep learning which is the concept of a GAN -- Generative Adversarial Network.

Maximizing the Dice loss (Eq.~\ref{eq:dice_loss}) forces the neural network to output a segmentation that is similar to the ground-truth one. However, in this loss function there is no preference for the transit signal to be necessarily periodic. Thus, the network might produce predictions that minimize the Dice loss but do not look `authentic', i.e., similar to real transits. An experienced exoplanet astronomer that would examine such an output would immediately be able to exclude an unauthentic segmentation. It is therefore required to add some kind of a penalizing mechanism to the network during training, so as to exclude those false predictions.

This problem is not unique to our setup only, but is common in the training of neural networks. There is a trade-off between minimizing the distortion (the Dice loss in our case) and the naturalness of the reconstructed signal (see for example the analysis for the case of super-resolution by \citet{BlaMic2018}). 
Thus, one may add a loss term in the training of the neural network that pushes the output distribution to resemble the true data distribution. GAN is a very popular strategy for achieving this goal.

A GAN comprises two neural networks, where one network (the `generator') generates `candidate' signals and the other (the `discriminator') evaluates them and discriminates between actual signals and ones produced by the generator. The training objective of the generator is to challenge the discriminator and increase its error rate (i.e. `fool' it by producing novel synthesised instances that appear to be genuine and cannot be distinguished from real data). The discriminator discriminates between genuine instances and artificial candidates produced by the generator \citep[e.g.][]{Gooetal2014}. 

Usually, GANs are used to generate purely new signals from some initial distribution. However, in our implementation, we use the generator to tag the samples that occur during transits, essentially producing a new sequence. The discriminator examines the resulting sequence and evaluates how realistic it is as a sequence of transit events.

GANs are known to suffer from training instability. In particular, a known problem in their training is `mode collapse', where the generated examples represent only a small fraction of the real distribution (e.g. the generator might generate always the same real-looking image). As a solution, a variant of GAN called Wasserstein GAN (WGAN) has been proposed in which the loss of the discriminator is set to be the Wasserstein distance (a measure of the distance between two probability distributions, also known as the Earth Mover’s Distance) leading to a more stable training \citep{Arjetal2017}. 

However, WGAN suffers from another problem, which is exploding of the gradient norm. As a solution to this problem, yet another variant of GAN has been proposed -- the WGAN gradient penalty (WGAN-GP), which penalizes the norm of the gradient of the discriminator network \citep{Guletal2017}. This method is known to perform even better, as it enables stable training of a wide variety of GAN architectures with almost no hyperparameter tuning.

In our case, we may train a discriminator that distinguishes between the output of the segmentation network (which acts as the generator) and the real ground-truth segmentation sequences. This use of the GAN framework to improve training of a given network (e.g. our segmentation network) is known as adding an adversarial loss, as the discriminator in this case is used as an `additional loss' in the training of the segmentation network. Specifically, in addition to maximizing the Dice coefficient, our segmentation network also aims at `fooling' the discriminator, which makes its output more similar to the ground-truth data.  

\section{Current Implementation} \label{sec:impl}

\begin{figure}
\centering
\includegraphics[trim=15pt 15pt 0 0, scale=0.2]{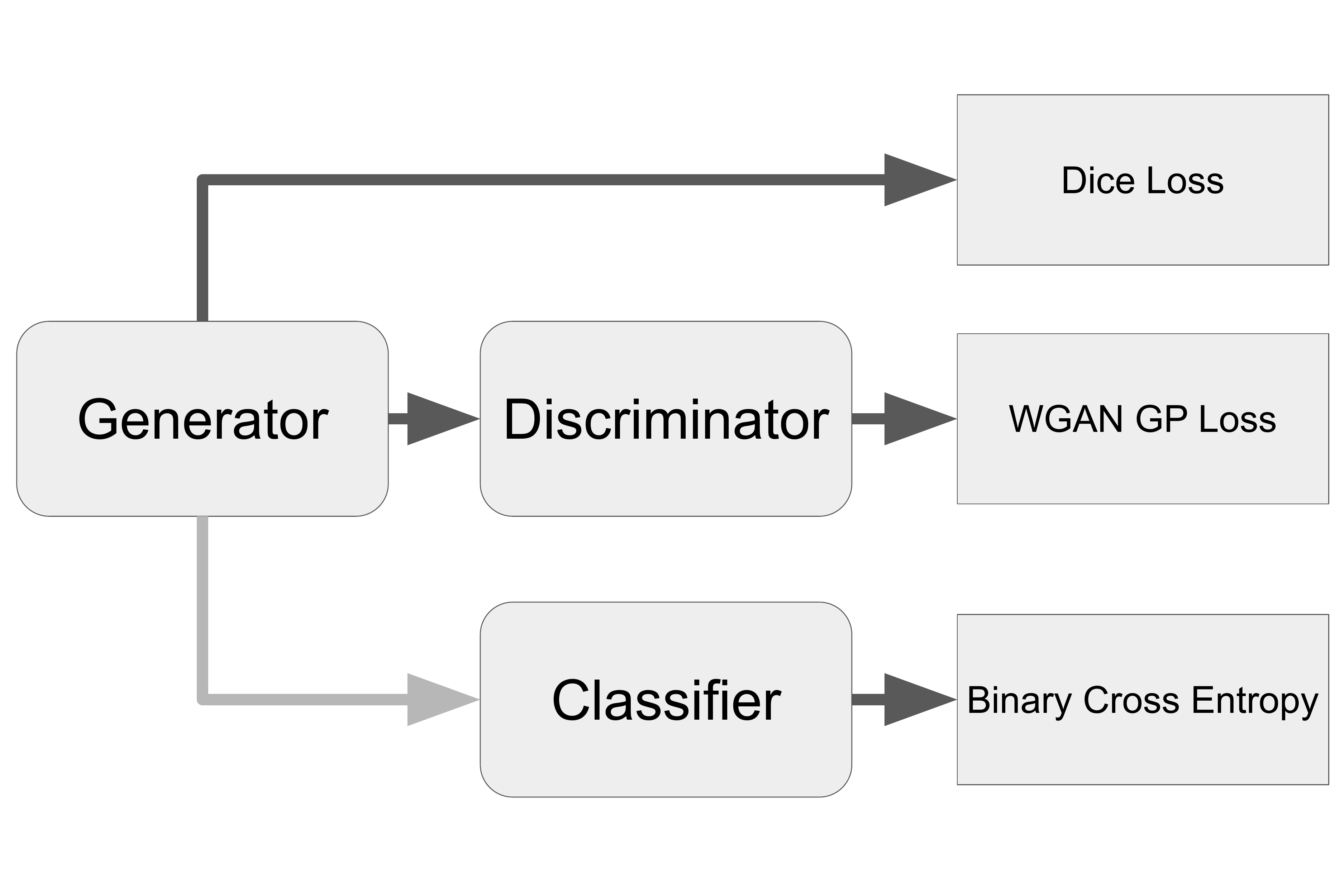}
\caption{A schematic illustration of the overall network structure. Note the gray arrow between the generator and classifier signifying no gradients flow through those connections.}
\label{fig:Structure}
\end{figure}

For the task of transit segmentation, we use the U-Net encoder-decoder architecture followed by the WGAN discriminator. In this setup, one may consider the U-Net as the generator of the WGAN network. Our model facilitates joint detection and segmentation with one architecture. Figure\ \ref{fig:Structure} presents the overall structure of this architecture. Notice that it contains a U-Net, a discriminator and a classification network. The gray arrow signifies that no gradients are flowing through the residual connections from the generator to the classifier, thus, only updating the classifier weights.

\begin{figure}
\centering
\includegraphics[trim=15pt 15pt 0 0, scale=0.48]{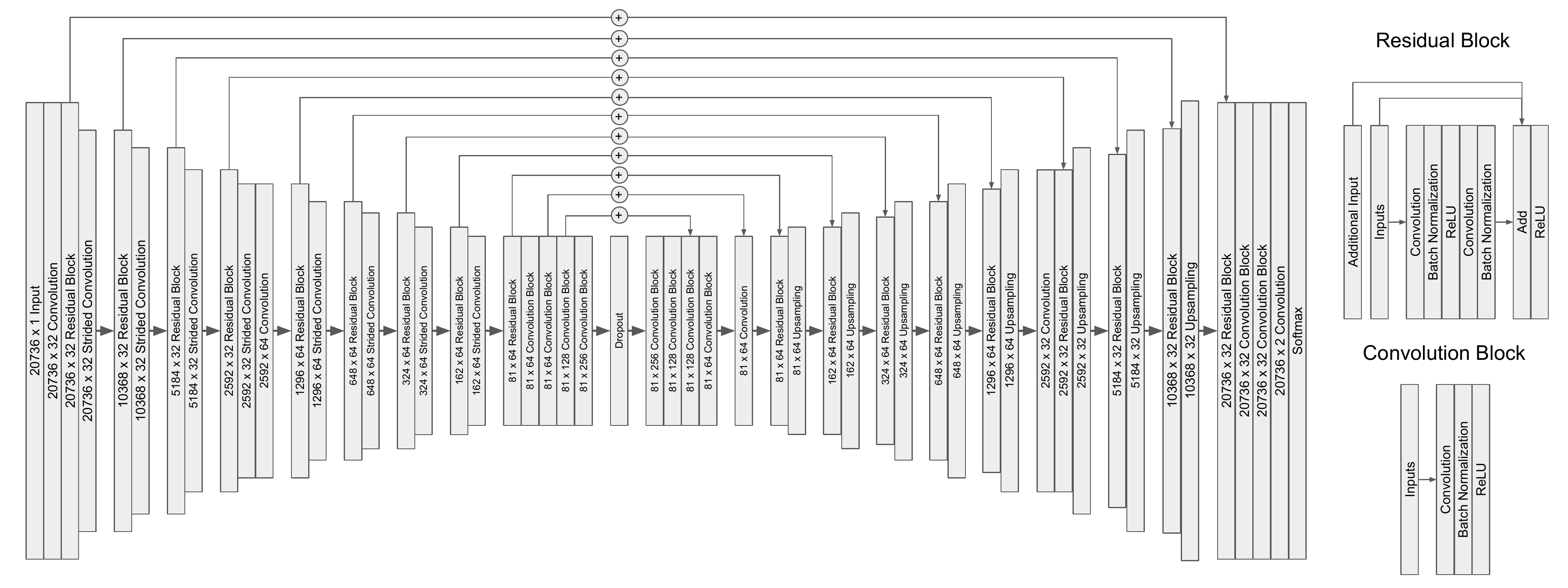}
\caption{A schematic illustration of the generator U-Net we use in our segmentation network. Note the residual "skip connections" between corresponding layers in the encoder and decoder.}
\label{fig:UNET}
\end{figure}

\begin{figure}
\centering
\includegraphics[trim=15pt 15pt 0 0, scale=0.48]{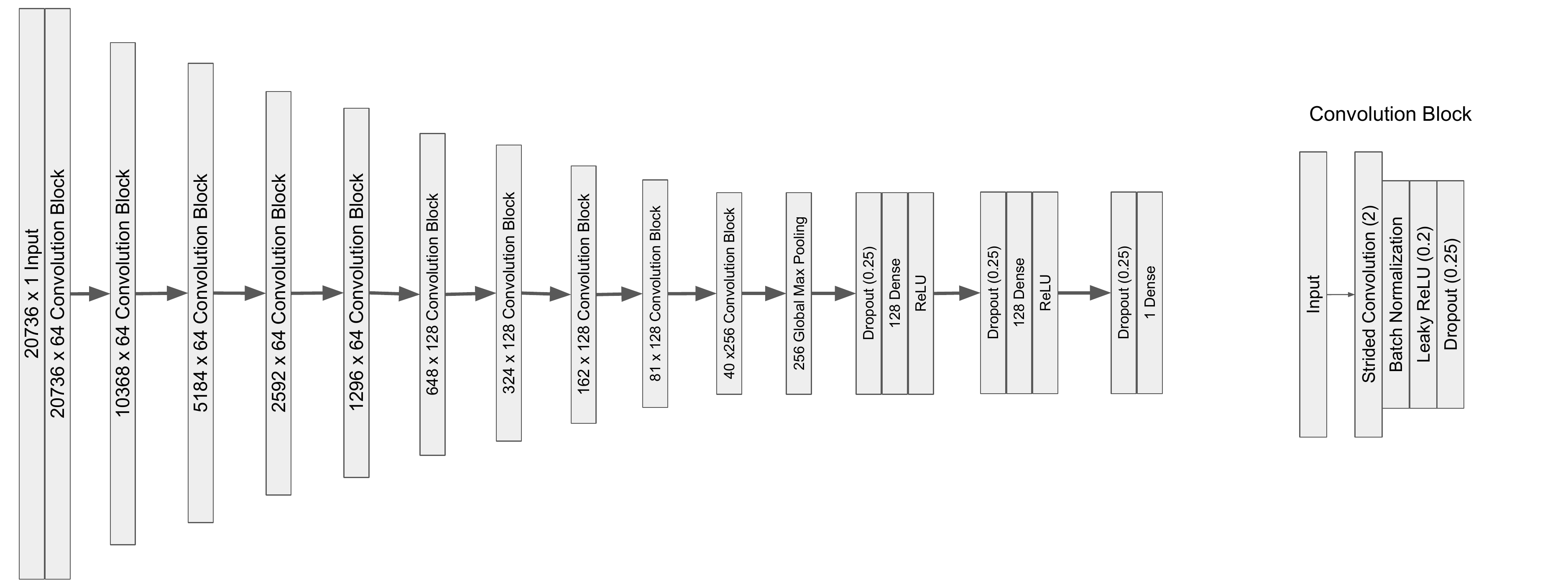}
\caption{A schematic illustration of the discriminator we use to produce the adversarial loss for training the segmentation network.}
\label{fig:Discriminator}
\end{figure}

\begin{figure}
\centering
\includegraphics[trim=15pt 15pt 0 0, scale=0.48]{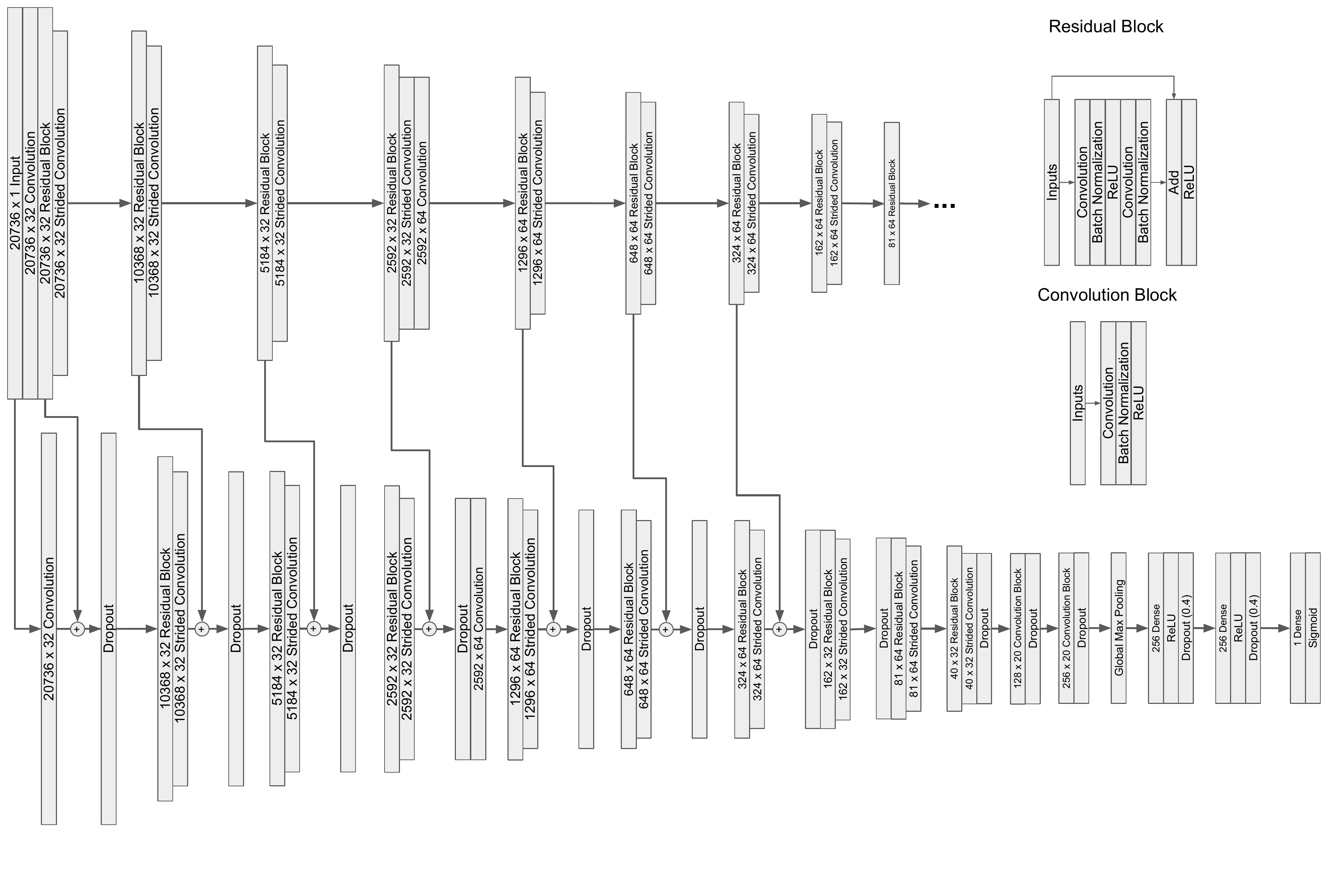}
\caption{A schematic illustration of the classifier network, which uses the residual connections from the generator U-Net.}
\label{fig:Classifier}
\end{figure}

The U-Net, the discriminator and the classification architectures are portrayed in Fig.~\ref{fig:UNET}, Fig.~\ref{fig:Discriminator}, and Fig.~\ref{fig:Classifier}, respectively. Note that the input dimension of the network is $20\,736$, which is a convient multiple of $2^8$, and longer than the original light curve, which was augmented by zero-padding.

The training process is split into two parts: (i) first, we train the generator and discriminator only on time series that contain transits, as a regular WGAN-GP, using the Dice loss combined with the adversarial loss, weighted $0.75$ and $0.25$ respectively when training the generator. In the second part of each training iteration, we freeze the weights of the generator and discriminator and train the classifier using binary cross-entropy on light curves, where only some of them contain transit signals. The fine technical details of the various architectures we used can be found in our code, which is publicly available on GitHub\footnote{\url{https://github.com/StrudelTAU/ShallowTransitsDL}} and archived in Zenodo \citep{Dvaetal2022}.

\section{Simulated data} \label{sec:sim}

We have used simulated data to train the network, and later to test and study its performance. 
The time sampling we have assumed for the simulations was the same time sampling introduced in the ETE-6 database by \citet{Jenetal2018}. The ETE-6 database had been released to the community to prepare for \textit{TESS} operations, and we therefore considered the sampling characteristics as representative of those of \textit{TESS}. Besides basing our time sampling on ETE-6 we did not use the ETE-6 data themselves. The main realistic feature which we found important in this sampling pattern was the inclusion of two sampling gaps attributed to downlink periods where the science operations of \textit{TESS} were assumed to be interrupted.

We assume that conventional methods (like the BLS) are sufficient in order to detect transits in the presence of white noise. We therefore focused our efforts and the simulated dataset on brighter stars, between magnitude $5$ and $10$, where the effects of red noise due to stellar activity are more significant (compared to white noise). 

The magnitude directly affects the uncorrelated photon noise, in a way that should depend on the characteristics of the observational apparatus. In order to estimate the white noise component in a two-minute cadence \textit{TESS} light curve, we approximated the curve in Figure 4 of \citet{Ricetal2015} by the following relation (assuming the brightness is quantified in the $I_\mathrm{C}$ band):
\begin{equation}
A_\mathrm{w} = \sqrt{108000 + 8670\,e^{0.94(I_\mathrm{C}-5)}}\ {\mu}\mathrm{mag}
\end{equation}

Following \citetalias{ZucGir2018}, we have used a Gaussian Process (GP) to simulate the noise, with a kernel comprising a squared-exponential component and a quasi-periodic one \citep[e.g.][]{Aigetal2016}. Combined with the white noise component, we got the following expression for the kernel of the noise GP:
\begin{equation}
k(t_i,t_j) = A_\mathrm{s}^2 \exp\left[-\left(\frac{t_i-t_j}{\lambda_\mathrm{s}}\right)^2\right]+A_\mathrm{q}^2 \exp \left[ -\frac{1}{2} \sin^2 \left(\frac{\pi \left(t_i-t_j\right))}{T_\mathrm{q}}\right)-\left(\frac{t_i-t_j}{\lambda_\mathrm{q}}\right)^2\right] +A_\mathrm{w}^2\delta\left(t_i-t_j\right)
\end{equation}
We assumed that the details of the red noise were related to stellar properties and not to the observational apparatus, and therefore we used the same distributions for the non-white-noise components as we had previously used in \citetalias{ZucGir2018}. 
Table \ref{table:hyper} summarizes the various hyperparameters of the GP that we have used.
Note that unlike the case in \citetalias{ZucGir2018} we have not added artificially any outlier samples to the noise.

\begin{deluxetable}{ccc}
\tablewidth{0pt}
\tablecaption{GP kernel hyper-parameter ranges \label{table:hyper}}
\tablehead{
\colhead{Hyper-parameter} & 
\colhead{Minimum value} & 
\colhead{Maximum value}
}
\startdata
$A_\mathrm{s}$ & $5\,\mu\,\mathrm{mag}$ & $125\,\mu\,\mathrm{mag}$ \\
$A_\mathrm{q}$ & $50\,\mu\,\mathrm{mag}$ & $125\,\mu\,\mathrm{mag}$ \\
$\lambda_\mathrm{s}$ &  $1\,\mathrm{min} $ & $10\,\mathrm{h}$ \\
$T_\mathrm{q}$ & $10\,\mathrm{h}$ & $500\,\mathrm{h}$ \\
$\lambda_\mathrm{q}$ &  $1\,000\,\mathrm{min} $ & $500\,\mathrm{h}$ \\
\enddata 
\end{deluxetable}

Unlike in \citetalias{ZucGir2018}, the ability to perform the segmentation might be affected by the detailed shape of the transits. Therefore, we could no longer settle for a simple trapezoid transit model. Instead we chose to use the publicly available code BATMAN which is capable of simulating transits quickly and accurately in a wide range of parameters and with various options to simulate the limb darkening \citep{Kre2015}. We chose to use a linear limb-darkening model \citep[e.g.][]{How2011}, with a single parameter $c_1$.

We drew a sample of stellar masses using a Salpeter Initial Mass Function between $0.3\,M_\sun$ and $2.0\,M_\sun$, which also provided us with the stellar radii, assuming a simplified mass-radius relation $(R/R_\sun) \simeq (M/M_\sun)$. We then drew the parameters of the planetary orbit: the period $P$, the planetary radius $R_\text{p}$ (in units of stellar radius), and the impact parameter $b$, which determined the orbital inclination. Table \ref{table:tran_parms} details the various distributions we used in order to draw all those parameters. The phase was drawn from a uniform distribution.

\begin{deluxetable}{cccc}
\tablewidth{0pt}
\tablecaption{Distributions of simulated transit parameters \label{table:tran_parms}}
\tablehead{
\colhead{Parameter} &
\colhead{Distribution} &
\colhead{Minimum value} &
\colhead{Maximum value} }
\startdata
$P$ & Log-Uniform & $2$ days & $9$ days \\
$R_\mathrm{p}/R_*$ & Log-Uniform& $1/100$ & $1/30$ \\
$b$ & Uniform & 0 & 1 \\
$c_1$ & Uniform & $0.5$ & $0.7$ \\
\enddata
\end{deluxetable}

\begin{figure}
\plotone{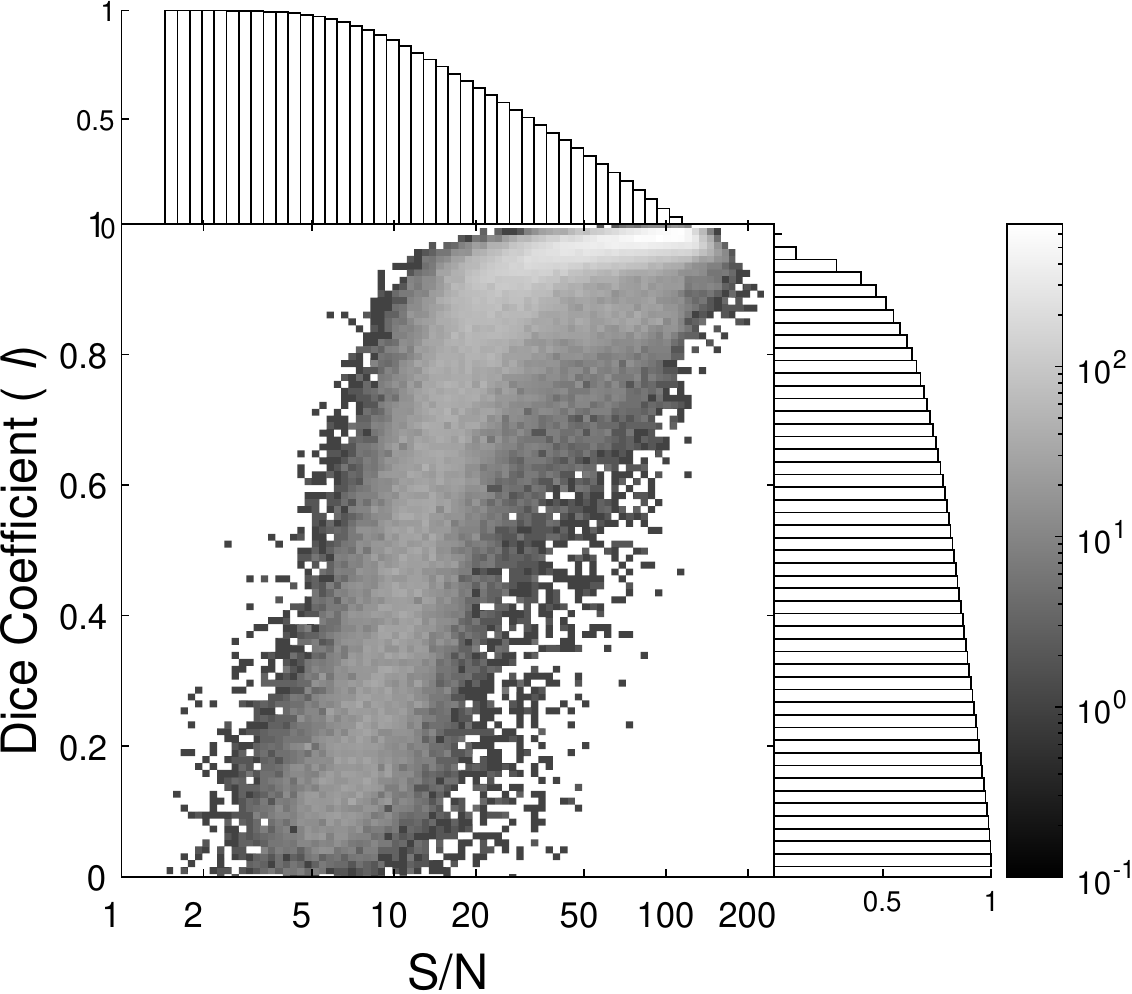}
\caption{Binned scatterplot of the obtained Dice coefficient as a function of the S/N for the $90\,000$ light curves containing transits. The greyscale coding of the scatterplot is visualized by the vertical bar on the right. The marginal distributions of the Dice coefficient and the S/N are also shown as normalized complementary cumulative histograms.}
\label{fig:SNR_Dice}
\end{figure}

In total, $100\,000$ light curves, each containing $20\,610$ samples, were simulated. For each of the generated light curves, we have injected a transit signal as described above. Thus, we eventually had a total of $200\,000$ light curves (consisting of pairs of which one contained a transit signal and the other did not). We have split the $100\,000$ pairs of light curves to $5\,000$ pairs for training, $5\,000$ pairs for validation (used mainly for hyper-parameter tuning), and $90\,000$ pairs for testing.
Note that simulating $100\,000$ light curves was a relatively easy process, and the computational burden was mainly related to the size of the training and validation sets, which is the reason for the larger size of the testing set.
The distribution of the S/N (as defined in \citealt{ZucGir2018}) of the $90\,000$ time series can be shown in Fig.~\ref{fig:SNR_Dice}, as the complementary cumulative distribution.

\section{Training} \label{sec:training}
We have trained the U-Net generator and the discriminator simultaneously using the $5\,000$ light curves that contained transits and their binary ground truth segmentation sequences together, while the classifier used both sets of light curves with and without added transits for $10\,000$ light curves in total. The global loss function contained coefficients that controlled the relative importance of segmentation (the Dice loss of the generator U-Net) and the evaluation of the discriminator (adversarial loss). The Dice coefficient was maximized during training, while the adversarial loss of the GAN discriminator was minimized using the Adam optimizer \citep{KinBa2014} with hyperparameters shown in Table \ref{table:hyperparams_gan} for $10\,000$ randomly chosen batches of size $32$ out of the $5\,000$ light curves mentioned above.

For the classification training, we used $10\,000$ light curves with and without added transits. 
The network was trained for $10\,000$ batches of $32$ data inputs each. This is approximately equivalent to $32$ epochs (i.e. going over the whole dataset $32$ times). Note that each batch is randomly selected from the whole dataset, using random permutation of the indices, thus guaranteeing going over all the data. After each batch of training with the generator-discriminator pair, a random batch of $32$ light curves (out of the $10\,000$ mentioned above) was used to train the classifier network using the Adam optimizer, with the hyperparameters shown in Table \ref{table:hyperparams_classifier}.

\begin{deluxetable}{cc}

\tablewidth{0pt}
\tablecaption{GAN Hyperparameters \label{table:hyperparams_gan}}
\tablehead{
\colhead{Parameter} &
\colhead{value} }
\startdata
learning-rate & $0.0001$ \\
beta\_1dropout & $0.54$ \\
beta\_2 & $0.9$ \\
batches & $10\,000$ \\
batch-size & $32$ \\
\enddata
\end{deluxetable}

\begin{deluxetable}{cc}
\tablewidth{0pt}
\tablecaption{Classifier Network Hyperparameters \label{table:hyperparams_classifier}}
\tablehead{
\colhead{Parameter} &
\colhead{value} }
\startdata
learning-rate & $0.001$ \\
dropout & $0.4$ \\
beta\_1 & $0.9$ \\
beta\_2 & $0.99$ \\
batches & $10\,000$ \\
batch-size & $32$ \\
\enddata
\end{deluxetable}

\section{Results} \label{sec:results}

The ability of a classifier network to identify light curves that contained transits was already demonstrated in \citetalias{ZucGir2018}. The main purpose of this work is to present a deep learning approach to identify the transit events in light curves  the light curves had already been labelled to contain transits.  Therefore, the results we present here are mainly examples that demonstrate the ability of the neural network to perform this task. 

\begin{figure}
\plotone{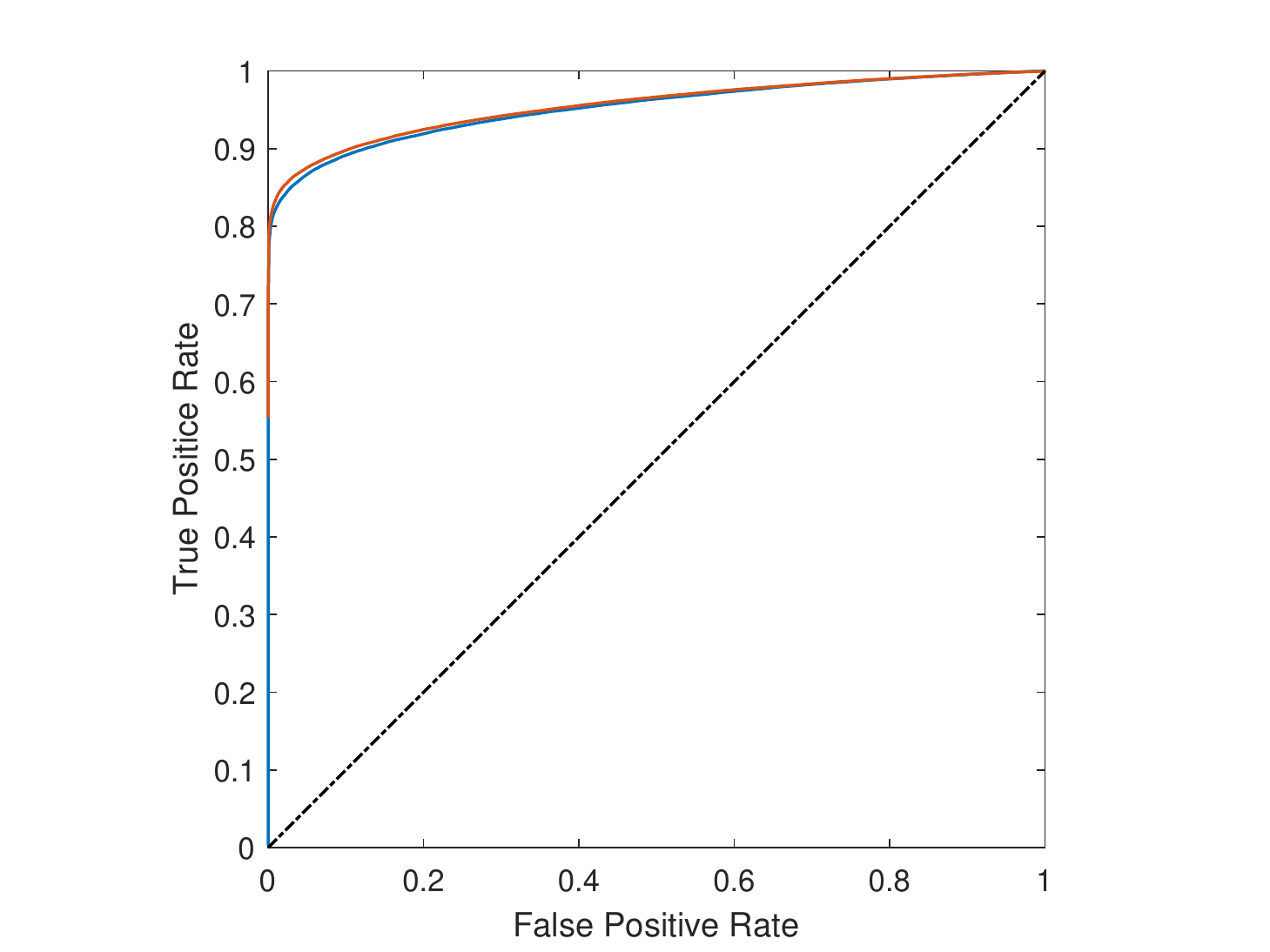}
\caption{The performance of the classifier network presented as a ROC curve. \textit{blue:} After training the classifier separately from the segmentation network; \textit{red:} after training the classifier together with the segmentation network. The dotted-dashed line is the so-called "no-discrimination" line, corresponding to randum guess.} 
\label{fig:ROC}
\end{figure}

However, we first show in Fig.~\ref{fig:ROC} that the classifier does perform satisfactorily in this context, in which it is trained together with the segmentation network. We show this using a Receiver Operating Characteristic (ROC) curve \citep[e.g.,][]{Faw2006}, which presents the true positive rate (TPR) as a function of the false positive rate (FPR). The blue curve shows the ROC for the classifier network, when trained alone, separately from the segmentation network. The red curve presents the results after training the two networks together. Clearly the classifier performance is only improving by this combined training, albeit slightly.

Fig.~\ref{fig:SNR_Dice} shows a binned scatterplot of the dependence of the Dice coefficient on the S/N of the light curves with transits. The marginal distributions of the Dice coefficient and the S/N are shown as normalized histograms representing the complementary cumulative distributions. Clearly, for many light curves the Dice coefficient is close to $1$. As can be expected, the performance generally tends to degrade for small S/N, and the performance below a S/N value of $10$ becomes quite poor. 

Most of the examples we chose to present here had a S/N between $10$ and $20$ and Dice coefficients higher than $0.9$. The last two examples are shown in order to sample S/N values beyond this restricted range. The characteristics of the examples are summarized in Table~\ref{table:examples}, including the assumed stellar magnitude of the star, the transit main parameters, the S/N and the obtained value of the Dice coefficient.

\begin{deluxetable}{ccccccccccccc}
\tablewidth{0pt}
\tablecolumns{13}
\tablecaption{Details of the presented examples \label{table:examples}}
\tablehead{
\colhead{} &
\colhead{} &
\multicolumn{3}{c}{Transit parameters} &
\multicolumn{6}{c}{GP parameters} &
\colhead{} &
\colhead{} \\
\hline
\colhead{} &
\colhead{} &
\colhead{period} &
\colhead{duration} &
\colhead{depth} &
\colhead{$A_\mathrm{w}$} &
\colhead{$A_\mathrm{s}$} &
\colhead{$A_\mathrm{q}$} &
\colhead{$\lambda_\mathrm{s}$} &
\colhead{$T_\mathrm{q}$} &
\colhead{$\lambda_\mathrm{q}$} \\
\colhead{} &
\colhead{magnitude} &
\colhead{(days)} &
\colhead{(hours)} &
\colhead{(mmag)} &
\colhead{($\mu$mag)} &
\colhead{($\mu$mag)} &
\colhead{($\mu$mag)} &
\colhead{(min)} &
\colhead{(h)} &
\colhead{(h)} &
\colhead{S/N} &
\colhead{Dice}
}
\startdata
A & $9.5$ & $6.69$ & $3.1$ & $1.10$ & $856$ & $26$ & $81$ & $13$ & $103$ & $85$ & $19.4$ & $0.908$ \\
B & $6.0$ & $4.74$ & $4.4$ & $0.22$ & $361$ & $95$ & $110$ & $3$ & $293$ & $65$ & $16.4$ & $0.901$ \\
C & $5.4$ & $3.61$ & $3.6$ & $0.24$ & $347$ & $55$ & $104$ & $285$ & $68$ & $71$ & $19.2$ & $0.904$ \\
D & $5.1$ & $6.26$ & $7.3$ & $0.25$ & $342$ & $7$ & $115$ & $335$ & $11$ & $25$ & $17.5$ & $0.931$ \\
E & $7.4$ & $4.67$ & $6.0$ & $0.30$ & $441$ & $32$ & $59$ & $7$ & $17$ & $212$ & $18.5$ & $0.917$ \\
F & $9.0$ & $7.28$ & $4.8$ & $0.61$ & $707$ & $80$ & $99$ & $13$ & $490$ & $58$ & $14.8$ & $0.907$ \\
G & $6.6$ & $4.49$ & $2.9$ & $0.23$ & $382$ & $42$ & $105$ & $13$ & $361$ & $157$ & $9.5$ & $0.873$ \\
H & $5.2$ & $6.54$ & $0.38$ & $0.59$ & $344$ & $117$ & $116$ & $36$ & $27$ & $30$ & $28.8$ & $0.029$
\enddata 
\end{deluxetable} 

\begin{figure}
\plotone{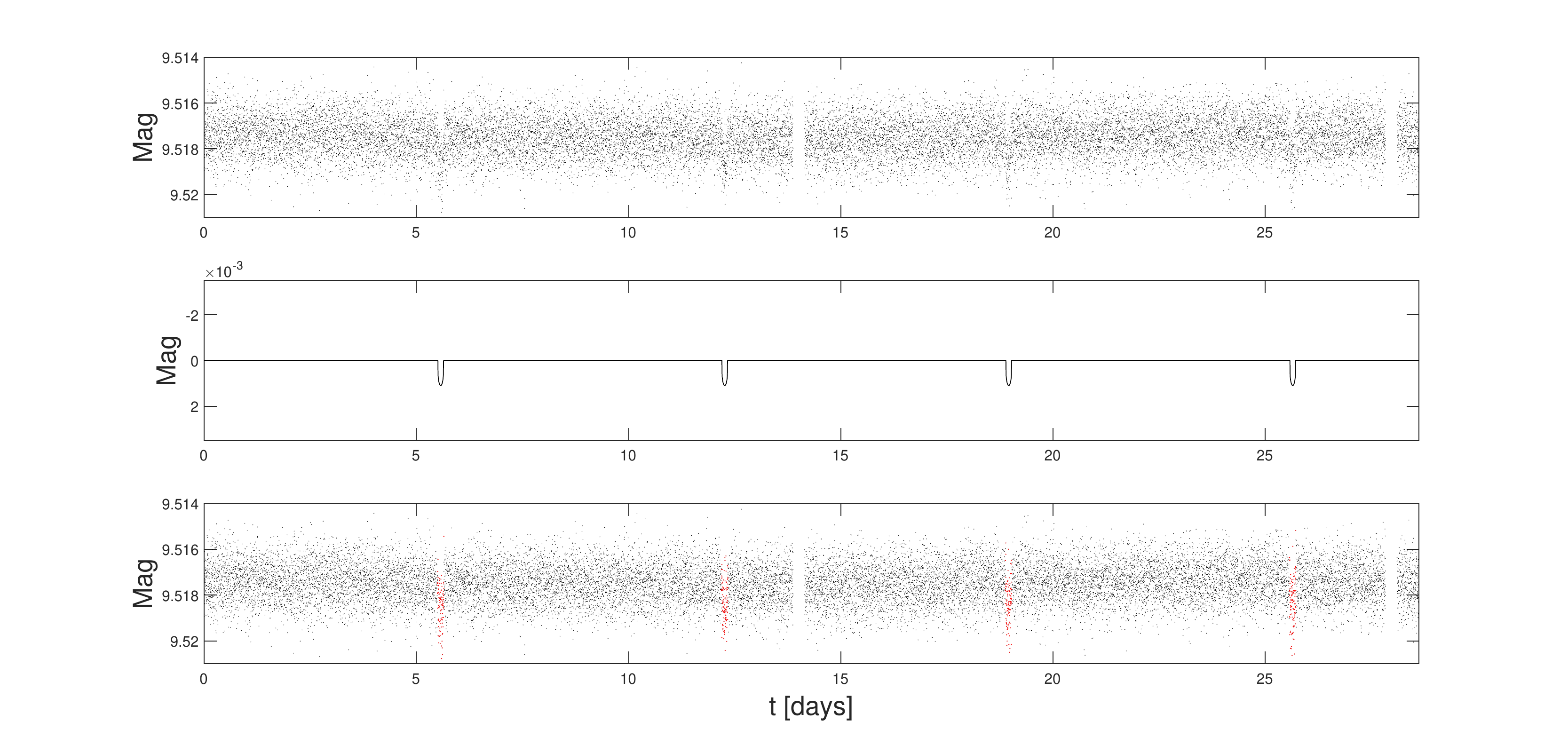}
\caption{Example A. The upper panel shows the input light curve. The middle panel shows the injected transit, while in the lower panel the light curve is shown again, but the samples that the network tagged as in-transit samples are marked in red. The y-axis scale is identical in all panels to facilitate visual comparison. Four transits can clearly be discerned.} 
\label{fig:example_1}
\end{figure}

Example~A is a simple and easy to study lightcurve. As can be seen in the top panel of Fig.~\ref{fig:example_1} there is no significant component of red noise -- there is almost no apparent long-term nor quasi-periodic variability, and the main source of noise is white noise. This is also evident in the parameters of the GP kernel (\ref{table:examples}). Note that this star is in the faint end of our simulated sample, which is indeed expected to be dominated by white noise. The transit is deep enough to that the transits can easily be spotted by eye in the light curve. The middle panel shows the injected transit signal at the same scale, while in the lower panel we highlighted by red the samples which the neural network identified as being in transit.

\begin{figure}
\plotone{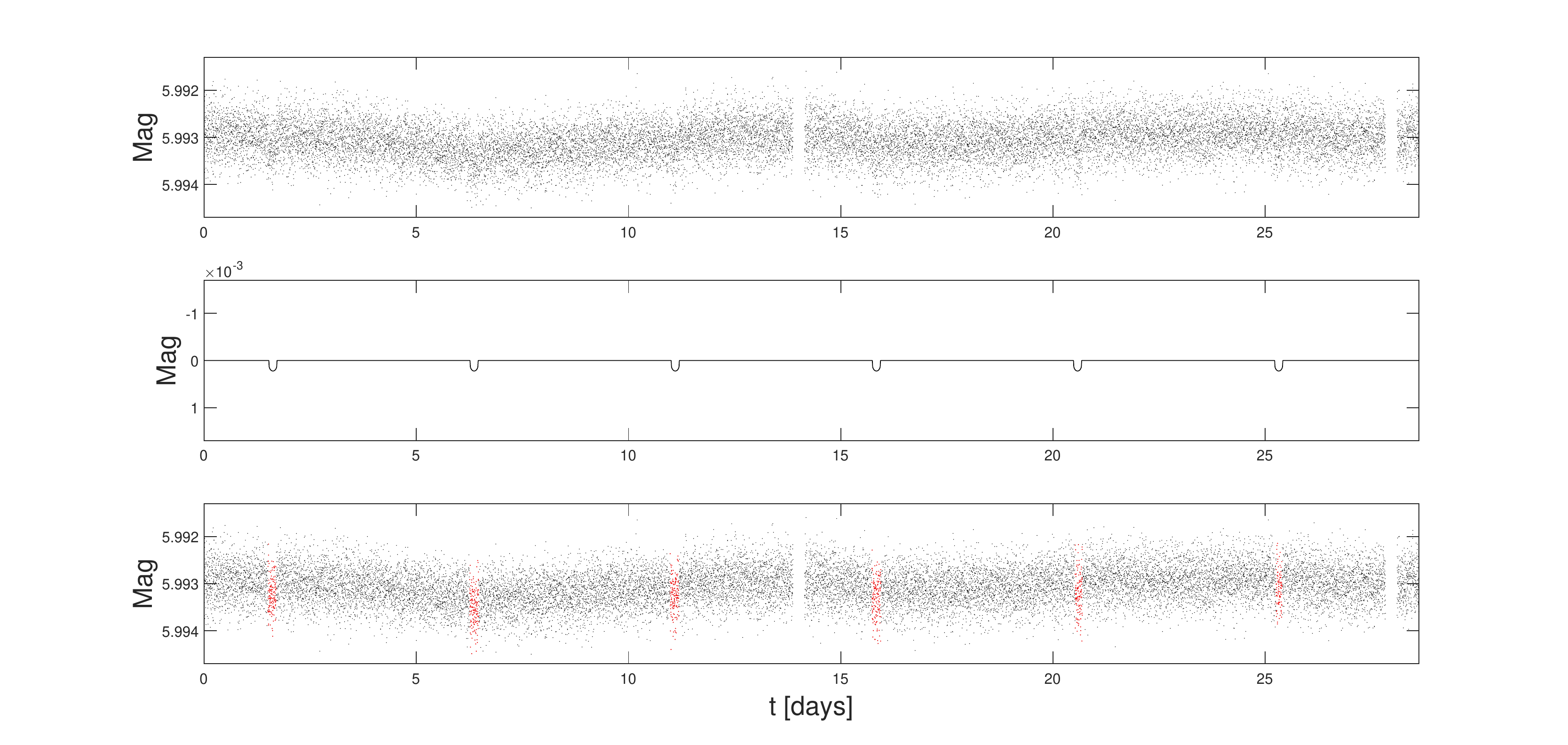}
\caption{Example B. Panels follow the structure of Fig.~\ref{fig:example_1}. Note the more prominent long-term variability in this specific case.} 
\label{fig:example_2}
\end{figure}

Example~B, shown in Fig.~\ref{fig:example_2}, is of a brighter sixth-magnitude star, and therefore the effect of red noise is supposed to be more significant. The transits are still visible when examining the light curve, but long-term red noise is also easily seen. Overall, the level of the noise is still low, which allows the very shallow transits to be discernible by eye.

\begin{figure}
\plotone{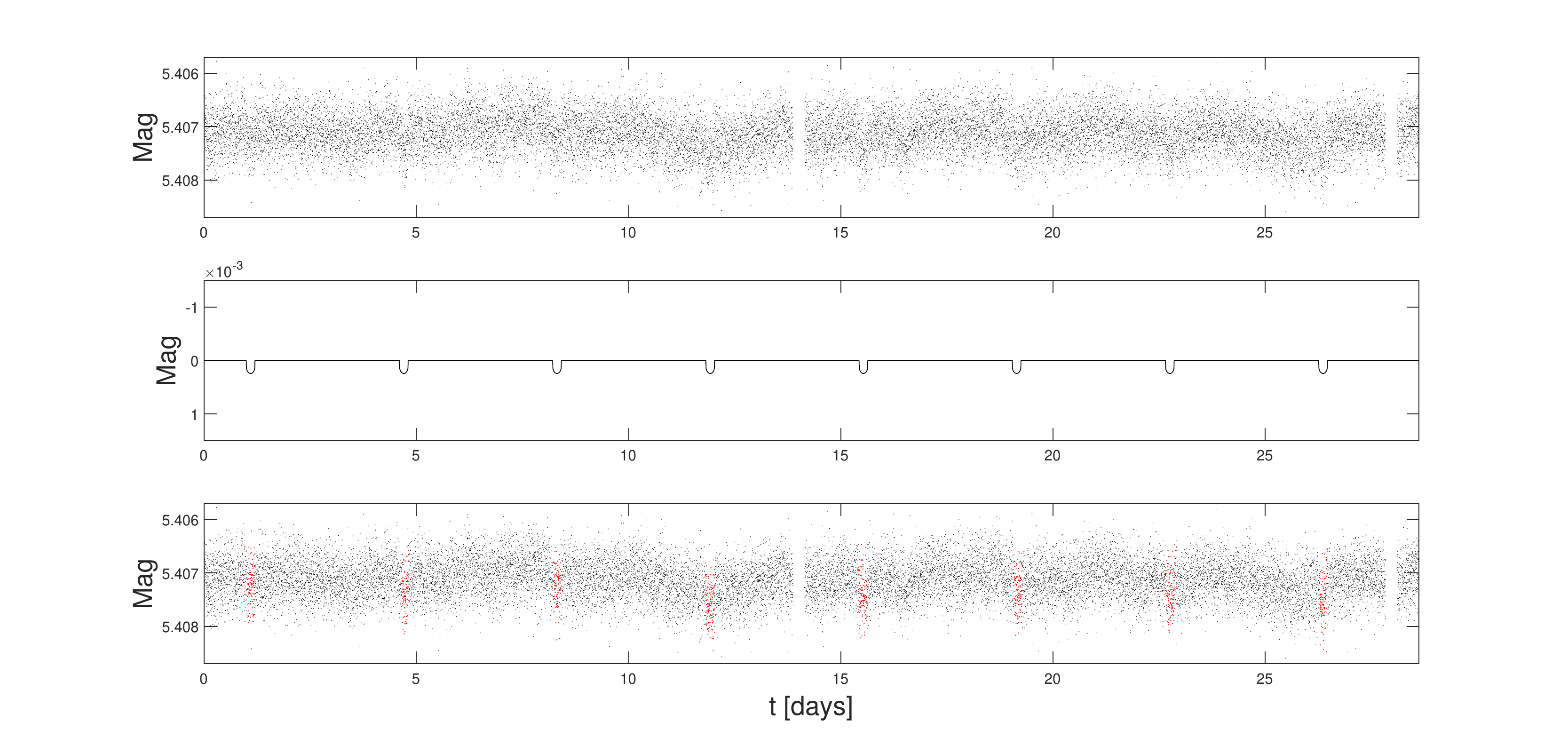}
\caption{Example C. Panels follow the structure of Fig.~\ref{fig:example_1}. Note the complex patterns of red-noise variability.} \label{fig:example_3}
\end{figure}

Fig.~\ref{fig:example_3} shows Example~C, where the red noise is more complicated.
The host star is even brighter than the previous example and one can see the many red-noise features in the light curve that are of similar amplitude as the transits. They are mainly caused by the relatively strong quasi-periodic term in the noise with a period of $68$ hours, which is close to the transit period of $3.61$ days (\ref{table:examples}). It is not easy to locate the transits by simple visual examination of the light curve.

\begin{figure}
\plotone{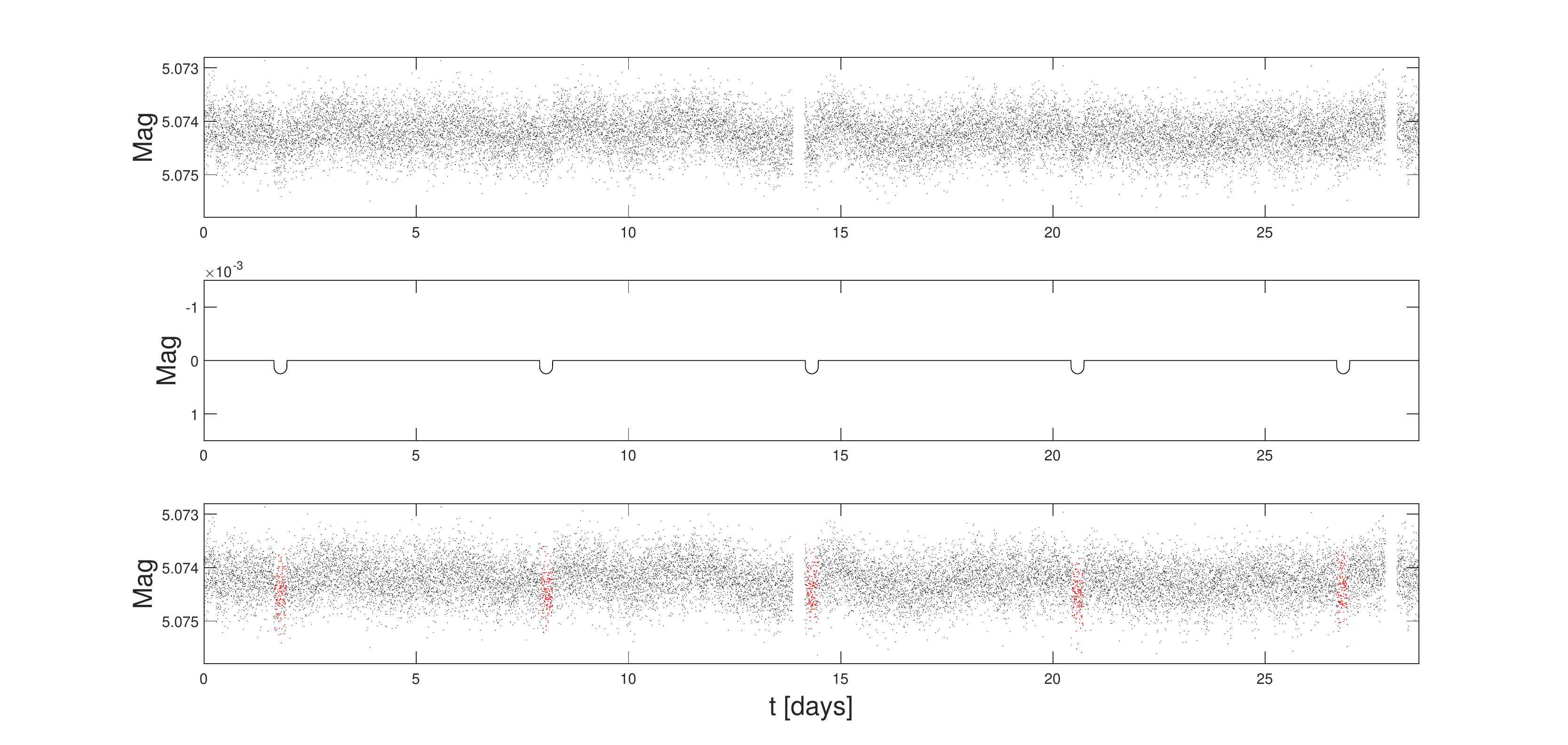}
\caption{Example D. Panels follow the structure of Fig.~\ref{fig:example_1}. Note the presence of many quasi-periodic transit-like features in the red noise.} 
\label{fig:example_4}
\end{figure}

The quasi-periodic component of the noise is even more prominent in Example~D shown in Fig.~\ref{fig:example_4}. The quasi-periodic noise creates many troughs in the light curve that can be easily mistaken by the human eye to be transits. The neural network still identifies correctly the timing of the periodic transits.

\begin{figure}
\plotone{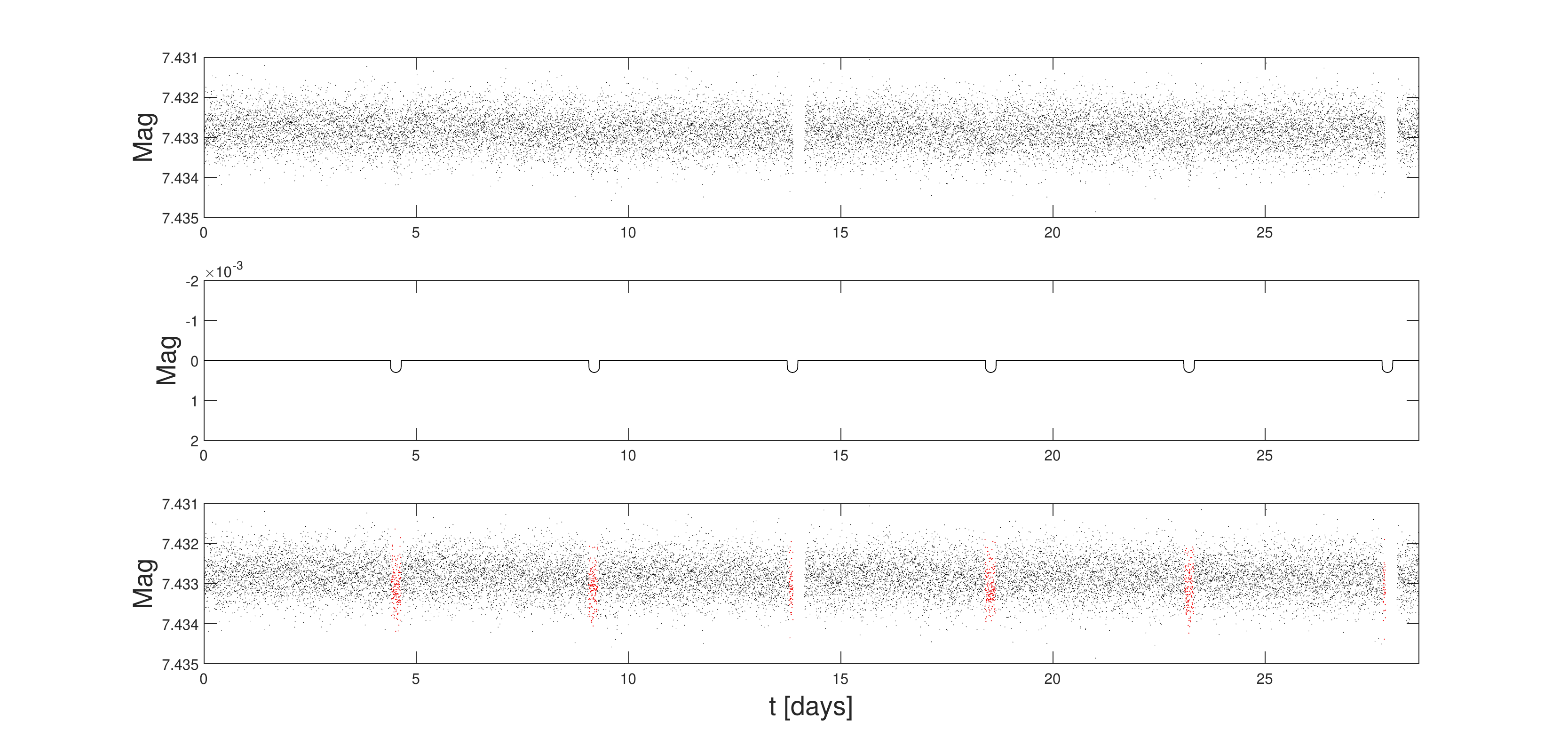}
\caption{Example E. Panels follow the structure of Fig.~\ref{fig:example_1}. Note the partial coalescence of the third and six transits with the downlink gap.} 
\label{fig:example_5}
\end{figure}

In Fig.~\ref{fig:example_5} we show Example E, which demonstrates the ability of the network to deal with the sampling gaps. In this example, both downlink gaps occur during transits of the exoplanet. The coincidence with the gaps does not hamper the ability of the network to perform a correct segmentation and identify the transit even in those times.

\begin{figure}
\plotone{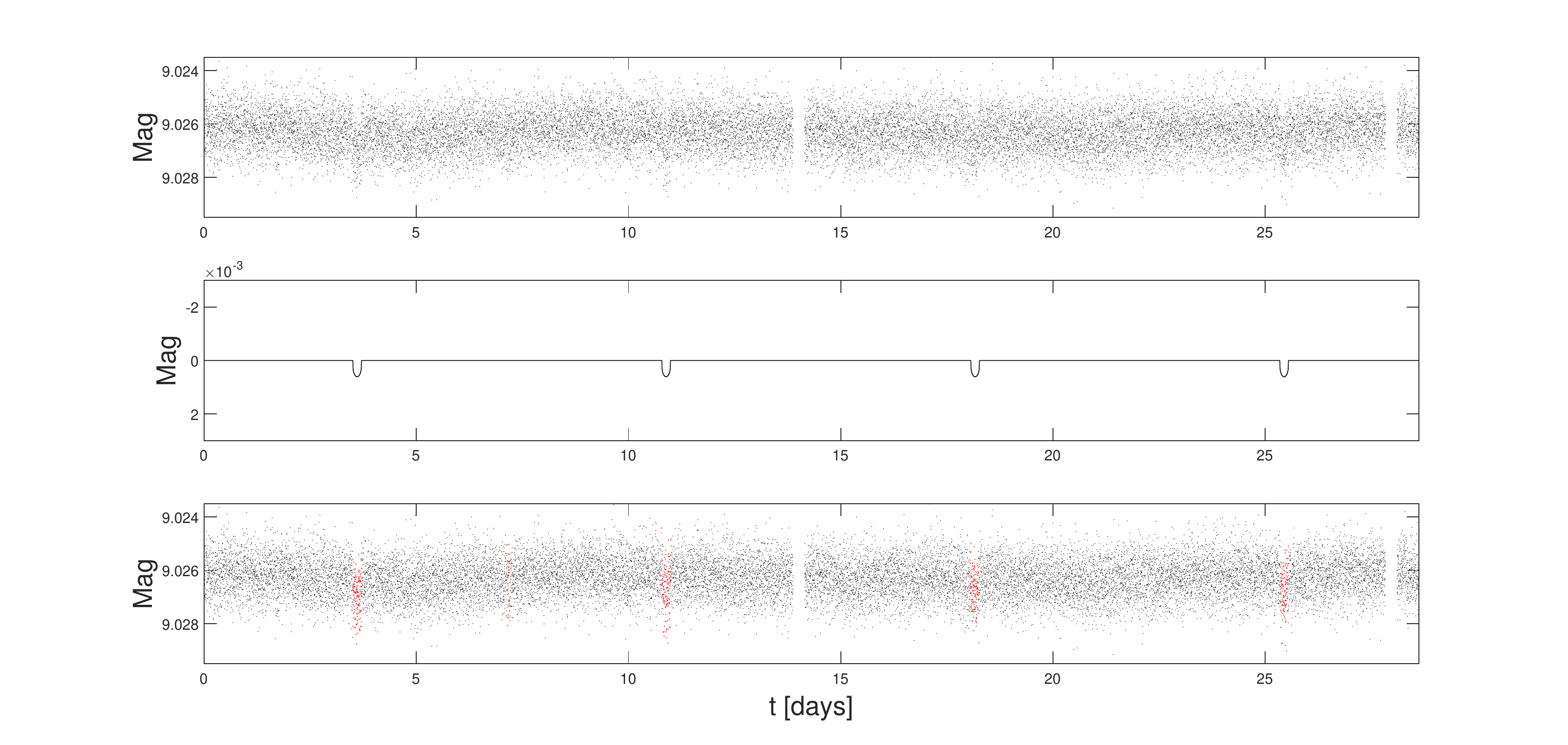}
\caption{Example F. Panels follow the structure of Fig.~\ref{fig:example_1}. Note the wrong identification of a short transit around day $7$.} 
\label{fig:example_6}
\end{figure}

Obviously, real life is never perfect, and errors are bound to occur especially for low S/N. Fig.~\ref{fig:example_6} shows Example~F, which is one of those rare events, where the network identified a transit when none existed. This light curve is dominated mainly by white noise, but with a large amplitude (the star is faint) making for a somewhat lower S/N. The network in this case wrongly identifies a very short transit at some point in day~$7$. This wrong identification of a transit event was not enough to lower considerably the Dice coefficient.

\begin{figure}
\plotone{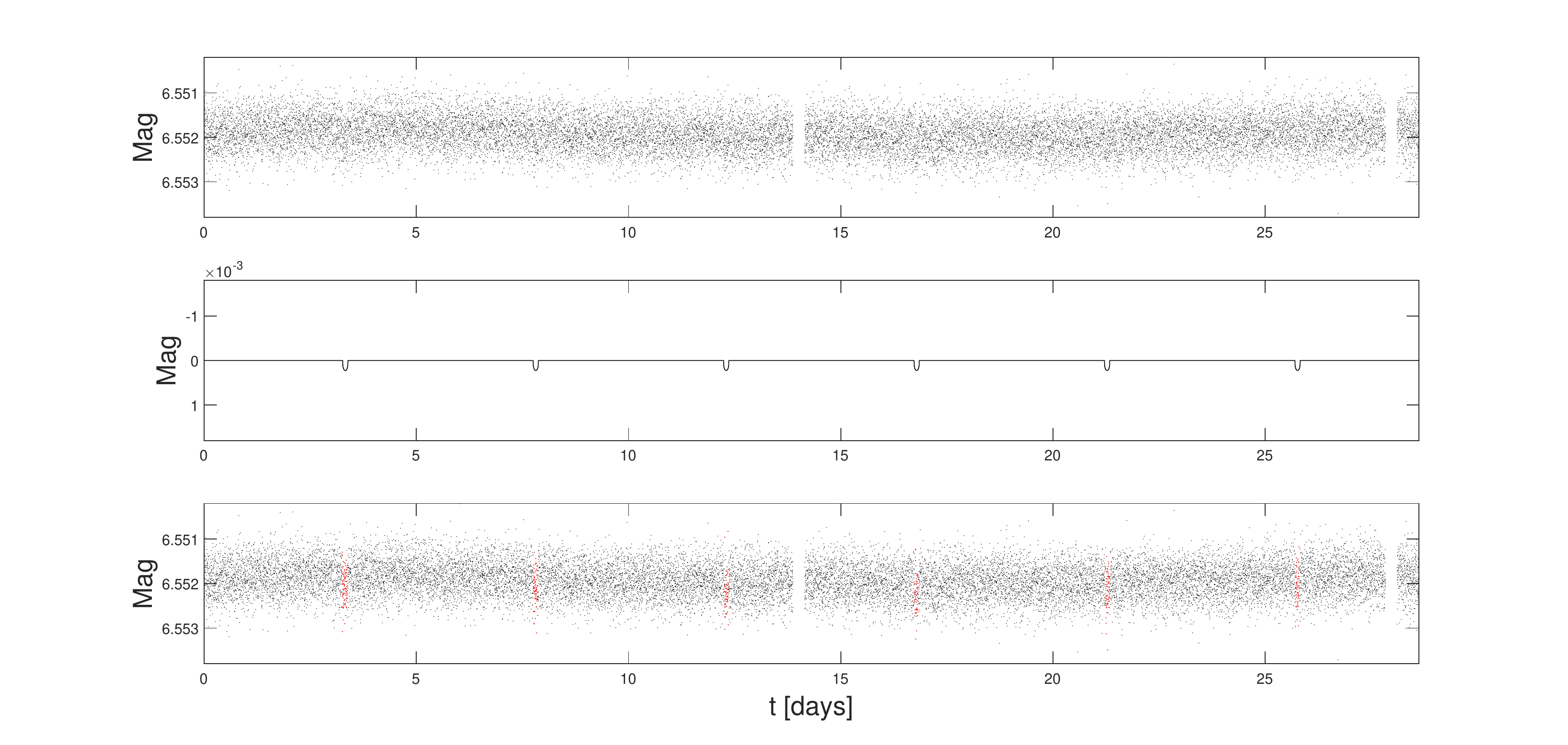}
\caption{Example G. Panels follow the structure of Fig.~\ref{fig:example_1}. All transits were identified in spite of the low S/N, resulting in a high Dice coefficient value.} 
\label{fig:example_7}
\end{figure}

\begin{figure}
\plotone{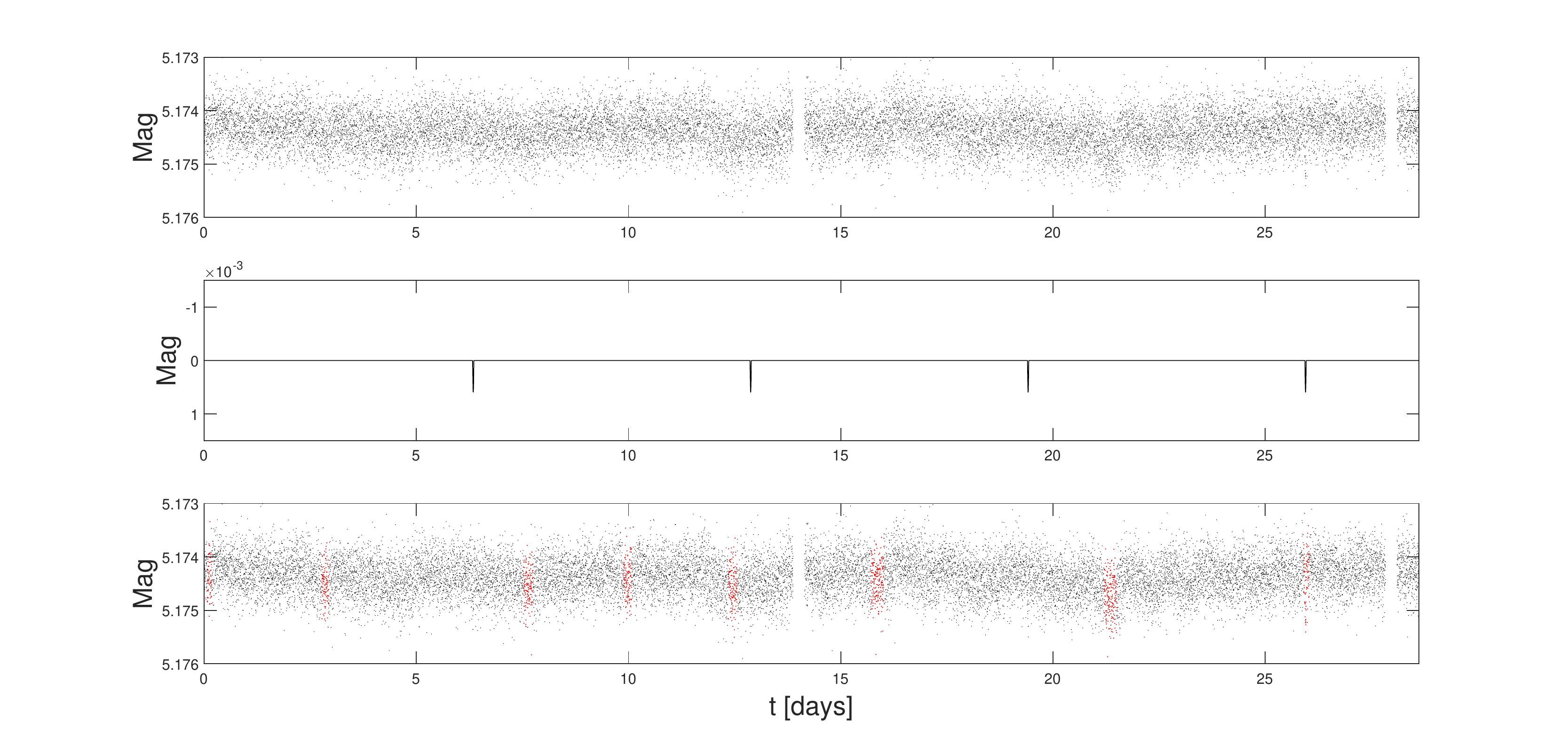}
\caption{Example H. Panels follow the structure of Fig.~\ref{fig:example_1}. The segmentation is completely wrong, except for the transit around day $26$, as reflected by the very low Dice coefficient.} 
\label{fig:example_8}
\end{figure}

The last two examples demonstrate cases of S/N values beyond the range between $10$ and $20$, and serve to substantiate the validity of the Dice coefficient as a diagnostic for the segmentation performance. Example~G, shown in Fig.~\ref{fig:example_7} shows a case of low S/N, below $10$, which nevertheless exhibits a relatively high Dice coefficient, almost $0.9$. This Dice value seems to be perfectly justified based on the satisfactory segmentation. On the other hand, Example~H, in Fig.~\ref{fig:example_8}, shows a case with a high S/N, above $20$, but very low Dice coefficient, in line with the very poor segmentation performance as seen in the figure. This poor performance is probably related to the very short transit duration, combined with a relatively strong presence of red and quai-periodic noise. 

\section{Discussion} \label{sec:discuss}

The approach we have described in this paper is meant to complement the detection neural network we have introduced in \citetalias{ZucGir2018}, where we have applied a neural network classifier to label simulated light curves which contained exoplanetary transits. Before any future analysis of the transits can be attempted, it is necessary to know when the transits actually take place. Only then can they be analyzed for their detailed shape, their precise timing, and most importantly, only then can one vet them and try to infer their nature -- whether they are genuinely exoplanetary transits or not. Following the terminology of computer vision, we have dubbed the task of labelling the in-transit samples `segmentation'. Both tasks, detection and segmentation, become extremely difficult when it comes to very shallow transits in the presence of red noise. We suggest that deep learning can significantly assist in performing those tasks, as we have tried to demonstrate with the simulations we presented.

The fact that the transits were periodic was very instrumental in using the GAN and the adversarial loss. Among other features of the transit signals, the network apparently learned to favour periodic signals. Once multiple transiting planets exist in the system, the challenge indeed becomes more complicated, but multiple aspects of periodicity still exist in the signal. If additional planets do not transit but only induce TTVs, we estimate that the effect on the segmentation network performance will not be that dramatic, but it still has to be verified. In our future work we intend to tackle those challenges (deviations from pure periodicity) as well.

For simplicity, we have employed in the training of the segmentation network a simple Dice loss that considers equally the in-transit and out-of-transit segments of the light curve. Assigning different weights to those areas may improve the performance and avoid the rare events of erroneous labels, like the one we show in our last example (Fig.~\ref{fig:example_6}). Furthermore, we note that the network usually reproduces quite well the phases of the transits, but the exact timing of the ingress and egress tend to suffer some more inaccuracies, probably caused by interfering features of the red noise. This might also be corrected by a more sophisticated loss function that would take into account the shape of the signal and the timing of the ingress/egress stages.

Unlike the network we have introduced in \citetalias{ZucGir2018}, the output of the network is not a binary decision regarding the presence of a transit signal, but a sequence of decisions tagging the original samples as in- or out-of-transit. The network itself outputs a sequence of real numbers, and a threshold is applied to those numbers to produce the binary segmentation sequence. In principle, a similar mechanism can be applied to perform detrending, in a way that would not destroy the transit signal. This will be another central aim of our future efforts -- detrend the signals and remove the red noise, without affecting the transit signal. This would be an extremely challenging task. 
  
We have trained the neural network such that it would tag transits that are really on the border of detection. The \textit{PLATO} mission will be targeting exactly such events. Deep learning techniques like the one we have presented here are bound to play a significant role in those efforts. Transits such as those presented in Figs.~\ref{fig:example_1} and \ref{fig:example_2} might be vetted reasonably well using current techniques. However, transits such as those shown in our next examples, the more difficult ones, might require additional observations, photometric, spectroscopic, or other that still have to be thought of.

In addition, similarly to \citetalias{ZucGir2018}, we added a binary classifier that uses information from the residual connections of the generator (Fig.~\ref{fig:Classifier}) to determine whether the light curve seems to contain a transiting planet signal, so that our final network performs the full task of identifying light curves with periodic transits and performing the segmentation.

Deep learning is an exploding field of data science. The exoplanetary community already acknowledges that and tries to use those techniques in its endeavor. However, most of the efforts focus around vetting transits using various variants of CNNs. The work we have presented here may serve as a reminder that there are other flavours of neural networks and training methods like U-Nets and GANs, which may also be very useful to exoplanet research.

\acknowledgments

This work was supported by a grant from the Tel Aviv University Center for AI and Data Science (TAD) and
by the Ministry of Science, Technology and Space, Israel.
R.G.\ acknowledges support by ERC-stg SPADE (grant No. 757497).


\begin{thebibliography}{}
\bibitem[Aigrain et al.(2016)]{Aigetal2016} Aigrain, S., Parviainen, G., \& Pope, B.~J.~S.\ 2016, \mnras, 459, 2408
\bibitem[Ansdell et al.(2018)]{Ansetal2018} Ansdell, M., Ioannou, Y., Osborn, H.~P., et al.\ 2018, \apjl, 869, L7
\bibitem[Arjovsky et al.(2017)]{Arjetal2017} Arjovsky, M., Chintala, S., \& Bottou, L. 2017, in Proceedings of the 34th International Conference on Machine Learning (ICML), 214
\bibitem[Blau \& Michaeli(2018)]{BlaMic2018} Blau, Y., \& Michaeli, T.\ 2018, in 2018 IEEE Conference on Computer Vision and Pattern Recognition (CVPR), 6228
\bibitem[Borucki et al.(2010)]{Boretal2010} Borucki, W.~J., Koch, D., Basri, G., et al.\ 2010, Science, 327, 977
\bibitem[Dattilo et al.(2019)]{Datetal2019} Dattilo, A., Vanderburg, A., Shallue, C.~J., et al.\ 2019, \aj, 157, 169
\bibitem[Deleuil et al.(2010)]{Deletal2010} Deleuil, M., Moutou, C., and the CoRoT Exoplanet Science Team 2010, in Physics and Astrophysics of Planetary Systems, ed. T.\ Montmerle, D.\ Ehrenreich and A.-M.\ Lagrange, EAS Pub. Ser. 41, 85
\bibitem[Dice(1945)]{Dic1945} Dice, L.~R.\ 1945, Ecology, 26, 297
\bibitem[Dvash et al.(2022)]{Dvaetal2022} Dvash, E., Peleg, Y., Zucker, S., \& Giryes, R.\ 2022, "StrudelTAU/ShallowTransitsDL", v1.0.0, Zenodo, doi:10.5281/zenodo.6304556
\bibitem[Fawcett(2006)]{Faw2006} Fawcett, T.\ 2006, PaReL, 27, 861
\bibitem[Goodfellow et al.(2016)]{Gooetal2016} Goodfellow, I.~J., Bengio, Y., \& Courville, A.\ 2016, Deep Learning (Cambridge, MA: MIT Press)
\bibitem[Goodfellow et al.(2014)]{Gooetal2014} Goodfellow, I.~J., Pouget-Abadie, J., Mirza, M., et al.\ 2014, in Proceedings of the 27th International Conference on Neural Information Processing Systems (NIPS 14), Vol. 2, ed. Z. Gharhamani et al. (Cambridge, MA: MIT Press), 2672 
\bibitem[Gulrajani et al.(2017)]{Guletal2017} Gulrajani, I., Ahmed, F., Arjovsky, M., Dumoulin, V., \& Courville, A.~C.\ 2017, in Advances in Neural Information Processing Systems (NeurIPS) 2017, 5767
\bibitem[He et al.(2016)]{Heetal2016}He, K., Zhang, X., Ren, S., \& Sun, J.\ 2016, in 2016 IEEE Conference on Computer Vision and Pattern Recognition (CVPR), 770
\bibitem[Howarth(2011)]{How2011} Howarth, I.~D.\ 2011, \mnras, 418, 1165 
\bibitem[Jenkins et al.(2018)]{Jenetal2018} Jenkins, J.~M., Tenenbaum, P., Caldwell, D.~A., et al.\ 2018, RNAAS, 2, 1
\bibitem[Kingma \& Ba(2014)]{KinBa2014} Kingma, D.~P., \& Ba, J.\ 2014, arXiv:1412.6980
\bibitem[Kov\'{a}cs et al.(2002)]{Kovetal2002} Kov\'{a}cs, G., Zucker, S., \& Mazeh, T.\ 2002, \aap, 391, 369
\bibitem[Kreidberg(2015)]{Kre2015} Kreidberg, L.\ 2015, \pasp, 127, 957
\bibitem[Lecun et al.(2015)]{Lecetal2015} LeCun, Y., Bengio, Y., \& Hinton, G.\ 2015, \nat, 521, 436
\bibitem[Lecun et al.(1998)]{Lecetal1998} LeCun, Y., Bottou, L., Bengio, Y., \& Haffner, P.\ 1998, IEEEP, 86, 2278
\bibitem[Liang et al.(2019)]{Liaetal2019} Liang, Y., Vanderburg, A., Huang, C., et al.\ 2019, \aj, 158, 1
\bibitem[Long et al.(2015)]{Lonetal2015} Long, J., Shelhamer, E., \& Darrell, T.\ 2015, in 2015 IEEE Conference on Computer Vision and Pattern Recognition (CVPR), 640
\bibitem[Milletary et al.(2016)]{Miletal2016} Milletary, F., Navab, N., \& Ahmadi, S.-A.\ 2016, in Proceedings of the 4th International Conference on 3D Vision (3DV), IEEE 2016, 565
\bibitem[Nair \& Hinton(2010)]{NaiHin2010} Nair, V., \& Hinton, G.~E.\ 2010,  in Proceedings of the 27th International Conference on Machine Learning (ICML-10), 807
\bibitem[Osborn et al.(2019)]{Osbetal2019} Osborn, H.~P., Ansdell, M., Ioannou, Y., et al.\ 2019, \aap, 633, A53
\bibitem[Rauer et al.(2016)]{Rauetal2016} Rauer, H., Aerts, C., Cabrera, J., et al.\ 2016, AN, 337, 961
\bibitem[Ricker et al.(2015)]{Ricetal2015} Ricker, G.~R., Winn, J.~N., Vanderspeck, R., et al.\ 2015, JATIS, 1, 014003
\bibitem[Ronneberger et al.(2015)]{Ronetal2015} Ronneberger, O., Philipp, F. \& Thomas, B.\ 2015,
in Medical Image Computing and Computer-Assisted Intervention (MICCAI 2015) part III, LNCS 9351, ed.\ N.~Navab et al.\ (Springer), 234
\bibitem[Rumelhart et al.(1986)]{Rumetal1986} Rumelhart, D.~E., Hinton, G.~E., \& Williams, R.~J.\ 1986, \nat, 323, 533
\bibitem[Schmidhuber(2015)]{Sch2015} Schmidhuber, J.\ 2015, NN, 61, 85
\bibitem[Shallue \& Vanderburg(2018)]{ShaVan2018} Shallue, C.~J., \& Vanderburg, A.\ 2018, \aj, 155, 94
\bibitem[S{\o}rensen(1948)]{Sor1948} S{\o}rensen, T.\ 1948, Biol.\ Skr.\ Dan.\ Vid.\ Sel., 5, 1
\bibitem[Zucker \& Giryes(2018)]{ZucGir2018} Zucker, S., \& Giryes, R.\ 2018, \aj, 155, 147

\end{thebibliography}
\end{document}